\documentclass[twoside,11pt]{article}

\usepackage{latexsym}
\usepackage{amssymb}
\usepackage{epsfig}

\makeatletter
\newenvironment{prog}{\vspace{1.0ex}\par
\setlength{\parindent}{3ex}
\setlength{\parskip}{0.0ex}
\obeylines\@vobeyspaces\tt}{\vspace{1.0ex}\noindent
}
\makeatother
\newcommand{\startprog}{\begin{prog}}
\newcommand{\stopprog}{\end{prog}\noindent}
\newcommand{\pr}[1]{\mbox{\tt #1}}   

\newcommand{\emptyline}{\vspace{1.0ex}} 
\newcommand{\ol}[1]{\overline{#1}}  
\newcommand{\sleq}{\leqslant}
\newcommand{\Xc}{{\cal X}}
\newtheorem{example}{Example}

\catcode`\_=\active
\let_=\sb
\catcode`\_=12
\makeatother

\begin{document}
\sloppy
\pagestyle{myheadings} 
\markboth{WLPE'01}{An Integrated Development Environment for Declarative
Programming} 

\title{An Integrated Development Environment for Declarative
Multi-Paradigm Programming\footnote{In A. Kusalik (ed),
Proceedings of the Eleventh Workshop on Logic
Programming Environments (WLPE'01),
December 1, 2001, Paphos, Cyprus. 
COmputer Research Repository (http://www.acm.org/corr/), cs.PL/0111039;
whole proceedings: cs.PL/0111042.
This research has been partially supported by the
German Research Council (DFG) under grant Ha 2457/1-2
and by the DAAD under the PROCOPE programme.}}
\author{Michael Hanus\thanks{
Institut f\"ur Informatik, Christian-Albrechts-Universit\"at Kiel,
Olshausenstr.~40, D-24098 Kiel, Germany, {\tt mh@informatik.uni-kiel.de}}\\CAU Kiel
\and
Johannes Koj\thanks{
Lehrstuhl f{\"ur} Informatik II, RWTH Aachen, Germany,
{\tt johannes.koj@sdm.de}}\\
RWTH Aachen
}

\date{}
\maketitle

\begin{abstract}
In this paper we present CIDER (Curry Integrated Development EnviRonment),
an analysis and programming environment for the
declarative multi-paradigm language Curry.
CIDER is a graphical environment to support the development
of Curry programs by providing integrated tools for the
analysis and visualization of programs.
CIDER is completely implemented in Curry using libraries
for GUI programming (based on Tcl/Tk) and meta-programming.
An important aspect of our environment
is the possible adaptation of the development environment
to other declarative source languages (e.g., Prolog or Haskell)
and the extensibility
w.r.t.\ new analysis methods. To support the latter feature,
the lazy evaluation strategy of the underlying implementation
language Curry becomes quite useful.
\end{abstract}

\section{Overview}
\label{sec-overview}

CIDER is a graphical programming and development environment for
the construction and debugging of declarative multi-paradigm programs.
Although the current implementation of CIDER is targeted at
the multi-paradigm programming language Curry \cite{Hanus97POPL},
the intension is to provide a development platform for both
functional and logic languages since Curry integrates
the most important features from functional programming
(nested expressions, lazy evaluation, higher-order functions),
logic programming (logical variables, partial data structures,
built-in search), and concurrent programming (concurrent evaluation
of expressions with synchronization on logical variables).
In particular, the implementation of CIDER is based
on an intermediate language to which functional, logic,
and also integrated functional logic programs can be compiled
(e.g., see \cite{AlbertHanusVidal00LPAR,AntoyHanus00FROCOS,%
AntoyHanusMasseySteiner01PPDP,Hortala-GonzalezUllan01}).
Thus, CIDER can be adapted to other declarative languages
provided that there exists a front end to compile programs
into this implementation-independent format (there exists
also an XML representation for this intermediate language,
see \cite{AntoyHanusMasseySteiner01PPDP}).

\begin{figure}[t]
\begin{center}
  \epsfig{file=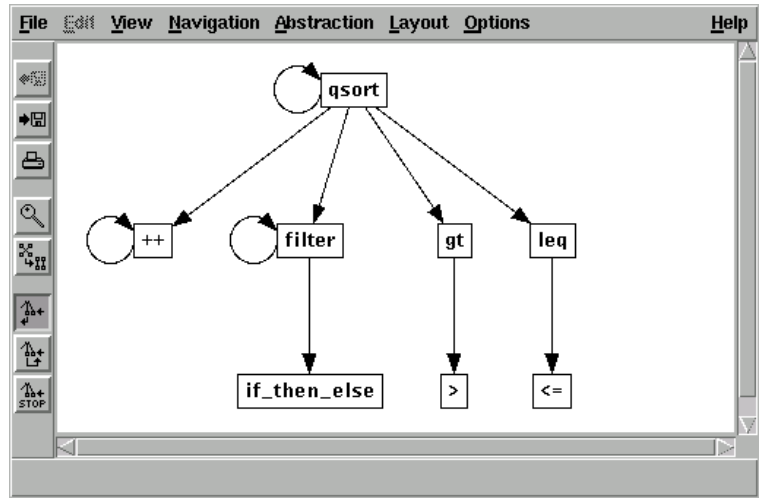,scale=0.55}
\end{center}\vspace{-3ex}
\caption{Visualization of a program dependency graph\label{fig-graph}}
\end{figure}

CIDER is an environment where various analysis and debugging
tools for declarative multi-paradigm languages are available.
Since the development of such tools is still an ongoing research,
CIDER is not designed as a closed system but it is 
intended as an open platform to integrate various
tools for analyzing and debugging programs.
Currently, CIDER consists of
\begin{itemize}
\item a program editor with the usual functionality,
\item various tools for analyzing properties of functions
defined in a program (types, overlapping definitions, complete definitions,
dependencies etc),
\item a tool for drawing dependency graphs,
\item a graphical debugger, i.e., a visualization of the evaluation
of expressions.
\end{itemize}
\begin{figure}[t]
\begin{center}
  \epsfig{file=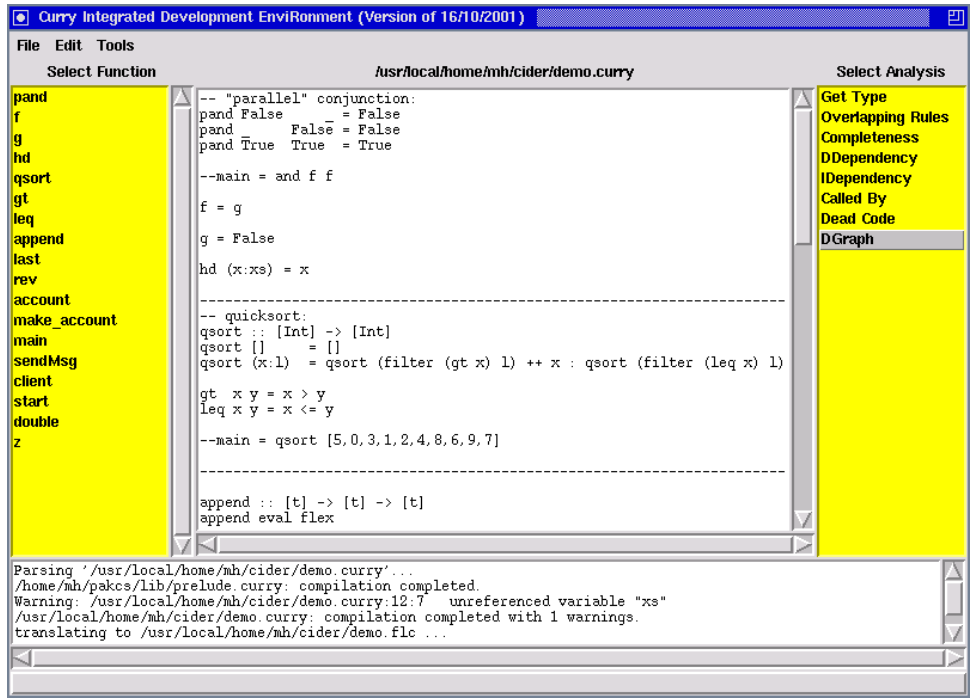,scale=0.45}
\end{center}\vspace{-3ex}
\caption{The main window of CIDER\label{fig-main}}
\end{figure}
To get an impression of the use of CIDER, Fig.~\ref{fig-main}
shows a snapshot after starting CIDER and loading a program.
The main window in the middle is an editor window for the current program.
On the left- and right-hand side, there is a list of the top-level functions
in the current file and a list of the currently available analysis tools
(see below for a description), respectively.
After selecting a function and an analysis
in the corresponding list boxes,
the function is analyzed and the analysis result is either shown in the bottom
window (if it is a textual result) or, if it is a graph,
it is visualized with the graph visualization tool
daVinci\footnote{\tt http://www.tzi.de/daVinci/}.
The current version of CIDER contains the following analysis tools
(which are useful but very simple and mainly included for
demonstration issues; see also Section~\ref{sec-ext}
for a description on how to add new analysis tools):
\begin{description}
\item[\pr{Get Type}:] Computes the function's type.
\item[\pr{Overlapping Rules}:] Shows whether the function is
defined by overlapping rules
(which might cause non-deterministic evaluations even for ground expressions).
This is interesting for logic programming but might be also useful
for purely functional programs.
\item[\pr{Completeness}:] Shows whether the function completely defined,
i.e., reducible on all ground constructor terms. Due to possible
overlapping rules, the current implementation is based
only on a sufficient criterion, i.e., the analysis results
are ``complete'' or ``might be incomplete''.
\item[\pr{(D/I)Dependency}:] Direct/indirect dependency, i.e.,
all functions that are directly or indirectly called in the rules
defining this function.
\item[\pr{Called By}:] Computes the list of all functions that call
this function in their defining rules.
\item[\pr{Dead Code}:] Computes the list of all top-level functions
in the currently loaded module that are not reachable
from the selected function.
\item[\pr{DGraph}:] Shows the dependency graph for the selected function.
This is a mixture as well as a graphical visualization of
\pr{(D/I)Dependency}, i.e., an arc is drawn from each function symbol
to all functions directly called in the rules defining this function
and all reachable function nodes are included in the graph.
\end{description}
For instance, the visualization computed by the analysis \pr{DGraph}
for the function \pr{qsort} (compare Fig.~\ref{fig-main}) is
shown in Fig.~\ref{fig-graph}.

Finally, CIDER contains also a graphical debugger/tracer
to visualize the evaluation of expressions.
Due to the fact that the operational semantics of the considered
language is important for this part of the programming environment,
we discuss it in Section~\ref{sec-debugger-impl}.

The rest of this paper is structured as follows.
The next section provides a short overview
of the main features of Curry as relevant for this paper.
Section~\ref{sec-impl} surveys the implementation of our
programming environment. Section~\ref{sec-ext}
sketches the necessary tasks to add new analysis tools to CIDER.
Finally, Section~\ref{sec-concl} contains our conclusions.

\section{Basic Elements of Curry}
\label{sec-intro-curry}

Although we mentioned above that our programming environment
can be adapted to other declarative languages than Curry,
the main motivation for the development of CIDER (and also Curry itself)
is to provide a common platform for declarative programming
where the most important declarative paradigms are smoothly integrated.
Therefore, Curry is our main target language
and we review in this section those elements of Curry
which are necessary to understand the functionality and
implementation of our programming environment.
More details about Curry's computation model and a complete
description of all language features can be found in
\cite{Hanus97POPL,Hanus00Curry}.

Curry is a modern multi-paradigm declarative language
combining in a seamless way features from functional,
logic, and concurrent programming
and supports programming-in-the-large with specific features
(types, modules, encapsulated search).
From a syntactic point of view, a Curry program is a functional
program extended by the possible inclusion of free (logical)
variables in conditions and right-hand sides of defining rules.
Curry has a Haskell-like syntax \cite{PetersonEtAl97},
i.e., (type) variables and function names usually
start with lowercase letters and the names of type and data constructors
start with an uppercase letter. The application of $f$
to $e$ is denoted by juxtaposition (``$f~e$'').

A Curry \emph{program} consists of the definition of functions
and the data types on which the functions operate.
Functions are evaluated in a lazy manner.
To provide the full power of logic programming,
functions can be called with partially instantiated arguments
and defined by conditional equations
with constraints in the conditions.
The behavior of function calls with free variables depends
on the evaluation annotations of functions which can be
either \emph{flexible} or \emph{rigid}.
Calls to rigid functions are suspended if a demanded argument,
i.e., an argument whose value is necessary to decide the applicability
of a rule, is uninstantiated (``\emph{residuation}'').
Calls to flexible functions are evaluated by a possibly
non-deterministic instantiation of the demanded arguments
to the required values in order to apply a rule (``\emph{narrowing}'').

\begin{example}\rm
\label{ex-conc}
The following Curry program defines the data types of
Boolean values and polymorphic lists (first two lines)
and functions for computing the concatenation of lists and the last
element of a list:
\startprog
data Bool   = True | False
data List a = []   | a : List a
\emptyline
conc :: [a] -> [a] -> [a]
conc eval flex       -- specify evaluation mode of conc as "flexible"
\vspace{0.7ex}
conc []     ys = ys
conc (x:xs) ys = x : conc xs ys
\emptyline
last xs | conc\,\,ys\,\,[x] =:= xs   = x   where x,ys free
\stopprog
The data type declarations define
\pr{True} and \pr{False} as the Boolean constants and
\pr{[]} (empty list) and \pr{:} (non-empty list) as the constructors for
polymorphic lists (\pr{a} is a type variable ranging over
all types and the type ``\pr{List\,\,a}'' is usually written as \pr{[a]}
for conformity with Haskell).

The (optional) type declaration (``\pr{::}'') of the function \pr{conc}
specifies that \pr{conc} takes two lists as input and produces
an output list, where all list elements are of the same
(unspecified) type.\footnote{Curry uses curried function types
where \pr{$\alpha$->$\beta$} denotes the type of all functions
mapping elements of type $\alpha$ into elements of type $\beta$.}
Since \pr{conc} is explicitly defined as
flexible\footnote{As a default,
all functions except for constraints are rigid.} (by ``\pr{eval flex}''),
the equation ``\pr{conc ys [x] =:= xs}'' can be solved by instantiating
the first argument \pr{ys} to the list \pr{xs} without the last argument,
i.e., the only solution to this equation satisfies that
\pr{x} is the last element of \pr{xs}.
\end{example}
In general, functions are defined
by (\emph{conditional}) \emph{rules} of the form
``$l \pr{~|\,} c \pr{~=\,} e \pr{~where~} vs \pr{~free}$''
where $l$ has the form
$f\,t_1 \ldots t_n$ with $f$ being a function, $t_1,\ldots,t_n$
data terms and each variable occurs only once,
the \emph{condition} $c$ is a constraint,
$e$ is a well-formed \emph{expression} which may also contain
function calls, lambda abstractions etc,
and $vs$ is the list of \emph{free variables} that
occur in $c$ and $e$ but not in $l$ (the condition and the \pr{where} parts
can be omitted if $c$ and $vs$ are empty, respectively).
The \pr{where} part can also contain further local function
definitions which are only visible in this rule.
A conditional rule can be applied if its left-hand side matches
the current call and its condition is satisfiable.
A \emph{constraint} is any expression of the
built-in type \pr{Success}.
Each Curry system provides at
least equational constraints of the form \pr{$e_1$\,=:=\,$e_2$}
which are satisfiable if
both sides $e_1$ and $e_2$ are reducible to unifiable data terms
(i.e., terms without defined function symbols).
However, specific Curry systems can also support more powerful
constraint structures, like arithmetic constraints on real numbers
or finite domain constraints,
as in the PAKCS implementation \cite{Hanus00PAKCS}.

Concurrent programming is supported by the concurrent conjunction
operator ``\pr{\&}'' on constraints, i.e., a non-primitive constraint
of the form ``\pr{$c_1$ \& $c_2$}'' is evaluated by solving
both constraints $c_1$ and $c_2$ concurrently. Since suspension
is controlled by the instantiation of arguments in calls to rigid
functions, concurrent computations are synchronized in a high-level
manner by logic variables as in concurrent constraint
programming \cite{Saraswat93}.

The operational semantics of Curry, precisely described in
\cite{Hanus97POPL,Hanus00Curry}, is a conservative extension
of lazy functional programming (if no free variables
occur in the program or the initial goal) and (concurrent) logic
programming. Since it is based on an optimal evaluation strategy
\cite{AntoyEchahedHanus00JACM}, Curry can be considered as a
generalization of concurrent constraint programming \cite{Saraswat93}
with a lazy (optimal) evaluation strategy.
Due to this generalization, Curry supports
a clear separation between the sequential (functional) parts of a program,
which are evaluated with an efficient and optimal
evaluation strategy, and the concurrent parts, based on the
concurrent evaluation of constraints, to coordinate
concurrent program units.

The concurrent conjunction operator ``\pr{\&}'' is a basic combinator
to create a fixed network of concurrent activities.
However, this primitive is too limited when a dynamically varying
number of processes with many-to-one communication structures
should be modeled, since this requires the merging of
message streams from different processes into a single message stream.
Doing that with a merger
function causes a set of problems as discussed in
\cite{Hanus99PPDP,JansonMonteliusHaridi93}.
Therefore, Janson et al.~\cite{JansonMonteliusHaridi93}
proposed the use of ports
for the concurrent logic language AKL which are generalized
in \cite{Hanus99PPDP} to support distributed programming in Curry.
In principle, a \emph{port} is a constraint between a multiset and a stream
which is satisfied if the multiset and the stream contain the same
elements (messages). In Curry a port is created by a constraint
``\pr{openPort p s}'' where \pr{p} and \pr{s} are free logical variables.
This constraint creates a multiset and a stream and combines them over a
port. Elements can be inserted into the multiset by sending them to
\pr{p} by the constraint ``\pr{send m p}''.
When a message is sent to \pr{p}, it will automatically be
added to the stream \pr{s} in order to satisfy the port constraint.
To support the implementation of distributed systems,
where the processes run on different machines,
ports can be also made externally accessible by assigning a
symbolic name to them. For instance, the I/O action\footnote{See
\cite{Wadler97} for a description of the monadic I/O concept of
Haskell which has been adapted without changes to Curry.}
\pr{(openNamedPort "$name$")} opens an external port with symbolic
name $name$ and returns the (infinite and lazy) stream of incoming messages.
If this port has been opened on machine $m$,
clients can access this port by executing
the I/O action \pr{(connectPort "$name$@$m$")}. This returns
a port $p$ to which they can send their messages. Note that
messages can also contain logical variables which provides
for a high-level mechanism to return values by instantiation
(rather than creating reply channels, see \cite{Hanus99PPDP}).
The port concept has been used to integrate object-oriented
features into Curry \cite{HanusHuchNiederau01IFL} which are often
necessary in GUI (Graphical User Interface) programming to keep the state
of user interfaces \cite{Hanus00PADL}.

\section{Implementation}
\label{sec-impl}

In this section we provide an overview on the implementation
of CIDER. This can be also seen as an example
to show that high-level declarative languages
provide the appropriate abstractions to implement
such advanced application in a modular and extensible way.
More details about the design and implementation of CIDER
can be found in \cite{Koj00}.

\subsection{Intermediate Representation Language}

In order to provide a high-level language for implementing
new analysis tools to be integrated in
our programming environment (see also Section~\ref{sec-ext}),
CIDER is completely implemented in Curry.
In order to implement program analyzers, one needs
a representation of programs as data objects.
For representing functional logic programs,
the direct representation of all program rules is not adequate
since the particular pattern-matching strategy is quite important
to reduce the search space and the length of derivations
\cite{AntoyEchahedHanus00JACM} and concrete languages often
differ in this strategy.
An appropriate data structure to describe such strategies
are definitional trees, introduced in \cite{Antoy92ALP}.
More recent approaches to the manipulation of
functional logic programs (e.g.,
\cite{AlbertHanusVidal00LPAR,AntoyHanusMasseySteiner01PPDP,%
Hortala-GonzalezUllan01}) advocate the explicit representation
of pattern matching by means of case constructs.

In the context of functional logic languages,
it is necessary to distinguish two kinds of 
case expressions in order to specify the flexible/rigid status
of functions. To be more precise,
we assume that all functions are defined by one 
rule whose left-hand side contains only pairwise different
variables as parameters and the 
right-hand side contains case expressions for pattern matching.
Thus, the basic syntax of programs in this representation
can be summarized as follows:\\[2ex]
{\small 
$
\begin{array}{lclllcll}
P & ::= & D_1 \ldots D_m & \mbox{ ~~~~  } &
e & ::= & v & \mbox{ (variable) } \\
D & ::= & f~v_1 \,\ldots\, v_n = e & \mbox{  } &
  & | & c~e_1\,\ldots\, e_n  & \mbox{ (constructor) } \\
&&&&  & | & f~e_1 \,\ldots\, e_n  & \mbox{ (function call) } \\
p & ::= & c~v_1 \,\ldots\, v_n  & \mbox{ } &
  & | & \mathit{case}~e_0~\mathit{of}~\{p_1\to e_1; \ldots; p_n \to e_n\}
                         & \mbox{ (rigid case) } \\
 &  & & &
  & | & \mathit{fcase}~e_0~\mathit{of}~\{p_1\to e_1; \ldots; p_n \to e_n\}
                         & \mbox{ (flexible case) } \\
 &  & & &
  & | & e_1~\mathit{or}~e_2 & \mbox{ (disjunction) } \\
\end{array}
$}\\[2ex]
where $P$ denotes a program, $D$ a function definition,
$p$ a pattern and $e$ an arbitrary expression.
A program $P$ consists of a sequence of
function definitions $D$ such that the left-hand side 
has pairwise different variable arguments and the right-hand side
is an expression $e$ composed by variables, constructors,
function calls, case expressions, and disjunctions.
A case expression has the form
$
\mathit{(f)case}~e~\mathit{of}~\{c_1~\ol{x_{n_1}} \to e_1,\ldots,c_k~\ol{x_{n_k}} \to 
e_k\}
$,
where $e$ is an expression, $c_1,\ldots,c_k$ are different 
constructors of the
type of $e$, and $e_1,\ldots, e_k$ are expressions.
The \emph{pattern variables} $\ol{x_{n_i}}$ are local
variables which occur only in the corresponding subexpression $e_i$.
The difference between $\mathit{case}$ and $\mathit{fcase}$ shows up when the
argument $e$ is a free variable:
$\mathit{case}$ suspends (which corresponds to residuation)
whereas $\mathit{fcase}$ nondeterministically binds this variable
to the pattern in a branch of the case expression
(which corresponds to narrowing).

\begin{example}
Consider the rules defining the (rigid) function ``$\sleq$'':
\[
\begin{array}{r@{~\sleq~}l@{~~=~~}l}
\tt 0    & \tt n & \tt True \\
\tt Succ~m & \tt 0 & \tt False \\
\tt Succ~m & \tt Succ~n & \tt m \sleq  n \\
\end{array}
\]
These rules can be represented by the following rule in our representation:
\[
\begin{array}{lllllll}
\tt x \sleq y = ~case~ x ~of & \tt \{0 & \tt \to & \tt True;\\
& \tt  ~Succ~x_1 & \tt \to&  \tt case~y~of~ & \tt \{0 \to ~ False; \\
&&&& \tt ~Succ~y_1 \to ~x_1 \sleq y_1 \}~ \}
\end{array}
\]
\end{example}
Based on this representation, some of the analyses
discussed in Section~\ref{sec-overview},
like the analysis for overlapping rules or completely defined
functions, can be implemented in a straightforward manner
by analyzing the structure of case expressions.
In order to cover all features of functional logic languages,
the definition of expressions can be extended by a few additional constructs,
like higher-order applications (``$\mathit{apply}~e_1~e_2$''),
partial applications, calls to external functions,
and existential quantification of variables
(e.g., see \cite{AlbertHanusVidal01FLOPS}).
Programs in this language, which is also called \emph{FlatCurry}\footnote{%
\tt http://www.informatik.uni-kiel.de/{\char126}curry/flat/},
can be simply represented as data objects
by a set of data type declarations similarly to
Example~\ref{ex-conc}. For instance, an entire program module
is represented as an expression of the type
\startprog
data Prog = Prog String [String] [TypeDecl] [FuncDecl]
                                 [OpDecl]   [Translation]
\stopprog
where the arguments of the data constructor \pr{Prog}
are the module name, the names of all imported modules,
the list of all type, function, and infix operator declarations
and a table to map external into internal names and vice versa.
Furthermore, a function declaration is represented as
\startprog
data FuncDecl = Func String Int TypeExpr Rule
\stopprog
where the arguments are the name, arity, type, and rule
(of the form ``\pr{Rule $\mathit{arguments}$ $\mathit{expr}$}'')
of the function (here we omit the other data type declarations).
The PAKCS implementation
of Curry \cite{Hanus00PAKCS} provides a library \pr{Flat}
for meta-programming which contains the definition
of such data types (also for representing data type declarations)
and an I/O action for reading a program file and translating
its contents into a \pr{Prog} term.

The FlatCurry representation of programs has been used
as an intermediate language to compile Curry
\cite{AntoyHanus00FROCOS,AntoyHanusMasseySteiner01PPDP}
or similar functional logic programs \cite{Hortala-GonzalezUllan01}
and to optimize declarative programs by partial evaluation
\cite{AlbertHanusVidal00LPAR,AlbertHanusVidal01FLOPS}.
However, it should be clear that this representation
is not restricted to Curry. Purely functional programs
can be translated into this intermediate language
without the use of $\mathit{fcase}$ and $\mathit{or}$ constructs.
Purely logic languages have only constraints as functions
and do not use the $\mathit{case}$ construct (although this
could be used for the translation of logic languages with coroutining).
For instance, the Prolog program
\startprog
app([],Ys,Ys).
app([X|Xs],Ys,[X|Zs]) :- app(Xs,Ys,Zs).
\stopprog
can be translated into the Curry program
\startprog
app [] ys zs = (ys =:= zs)
app (x:xs) ys (z:zs) = (x =:= z) \&> (app xs ys zs)
\stopprog
where ``\pr{\&>}'' denotes the sequential (``left-to-right'') conjunction
of constraints (replacing ``\pr{\&>}'' by the concurrent conjunction
``\pr{\&}'' corresponds to a logic program with coroutining).
Note that the introduction of auxiliary variables and explicit
equality constraints is necessary due to the left-linearity requirement
in Curry. The latter rules can be translated into the following
FlatCurry definition:
\begin{tabbing}
~~\pr{app xs ys zs} $=$\\
~~~~~~~~~~~~~~~~\pr{fcase xs of}~\{ \= \pr{[]}~~~~\= $\to$~ \= \pr{ys =:= zs};\\
\> \pr{x:x1} \> $\to$ \> \pr{fcase~zs~of}\\
\> \> \> ~~~~\{ \= \pr{z:z1} $\to$ \pr{(x =:= z) \&> (app x1 ys z1)} \}~\}
\end{tabbing}

\subsection{Program Analysis}
\label{sec-ana-impl}

As shown in Section~\ref{sec-overview}, our programming environment
is intended to integrate various analysis tools for declarative
programs. This demands for a unique interface for the implementation
of each program analysis to be integrated into CIDER.
Since the intermediate language FlatCurry is a reasonable
basis for writing program analyzers, we require that each
program analyzer must implement a function of type
\startprog
type ProgAnalysis = Prog -> [(String,AnaRes)]
\stopprog
where
\startprog
data AnaRes = Message String | Graph DvGraph
\stopprog
is the result type of analyzing an individual function
in a program.
Thus, a program analysis takes a program as input (\pr{Prog})
and produces a list of analysis results, where each analysis result
is a pair consisting of a function name and the associated
result of analyzing this function.
In our current implementation, it is sufficient that this
result is either a string  $s$ (``\pr{Message $s$}'', e.g., the
type of a function, ``overlapping''/``not overlapping'',
the list of called functions) to be shown in the bottom window
of the main interface, or a graph $g$ (``\pr{Graph $g$}'', e.g.,
the function's dependency graph) to be visualized in separate
window with the graph visualization tool daVinci
(see Fig.~\ref{fig-graph}).

As a simple example, the analysis of overlapping rules
can be implemented as follows:
\startprog
analyseOverlappings :: Prog -> [(String,AnaRes)]
analyseOverlappings (Prog _ _ _ funs _ _) = map overlapFun funs
  where overlapFun (Func name _ _ (Rule _ e))
              | orInExpr e = (name, Message "overlapping")
              | otherwise  = (name, Message "not overlapping")
\stopprog
where the function \pr{orInExpr} checks for occurrences of disjunctions
in an expression.

It is interesting to note that the lazy evaluation strategy
of our implementation language Curry becomes quite handy here.
Although a program analysis is defined as a function operating
on the entire program, which may cause a complex computation,
the lazy evaluation strategy performs the program analysis
in a demand-driven manner. If a user selects an analysis
(in the right window of the main interface, see
Fig.~\ref{fig-main}), the corresponding analysis function
is applied to the currently loaded program module.
Due to lazy evaluation, this application is not evaluated
at this analysis selection time but only if the user selects an
individual function in the left window of the main interface.
If such a function is selected, the analysis is performed
to show its result. However, note that only the selected function
is analyzed which can be done locally (e.g., ``overlapping'' analysis)
or might require the consideration of other parts of the program
(e.g., computing the dependency graph).
Furthermore, if the same function is selected again, the result is
directly available due the fact that lazy evaluation
evaluates each expression at most once.
This behavior is very desirable for our environment.
In a strict or imperative language, one needs some
additional effort to implement such a behavior.

\subsection{Main Interface}

The main graphical user interface (see Fig.~\ref{fig-main})
is implemented with the library
\pr{Tk} which supports a high-level implementation of GUIs
in Curry by exploiting the integrated functional and logic features
of the language \cite{Hanus00PADL}.
With this library, the structure of the interface
(i.e., the different widgets) is described as a data term
containing call-back functions for implementing
the functionality of the individual widgets.

One difficulty in the implementation of user interfaces
in a declarative language is the handling of internal states.
This is necessary for keeping the name of the currently
loaded program module, the selected program analysis,
the selected function etc.
GUI libraries for purely functional languages
(e.g., \cite{ClaessenVullinghsMeijer97ICFP})
advocates the use of monads for this purpose.
Since our GUI library is not based on monads but exploits
the functional and logic features of Curry,
we handle the internal state of the GUI in an object-oriented
style by exploiting Curry's port concept sketched in
Section~\ref{sec-intro-curry}.
Thus, the state of the main interface is kept as an argument
of a recursive function implementing a ``\pr{GUIServer}''.
Basically, this function has type
\startprog
serveGUI :: GUIServerState -> [GUIServerMsg] -> Success
\stopprog
where \pr{GUIServerState} contains all data in the
state of the main interface and
\pr{GUIServerMsg} are the messages to be processed
by the server. This function is implemented
by a case distinction on the different first messages
where the function is (tail recursively) called
with a modified state (depending on the message) and the list
of remaining messages.
For instance, the message \pr{Terminate} terminates the server
and the message ``\pr{SetAna $a$}'' sets the current program analysis.
Thus, we have the following defining rules, among others
(the function \pr{changeAna}
computes a new state with a changed current analysis component):
\startprog
serveGUI _     (Terminate:_)     = success \,--\,the end, no recursive call
serveGUI state (SetAna a : msgs) = serveGUI (changeAna a state) msgs
\stopprog
Now, the main interface is implemented as a GUI (based on the \pr{Tk}
library as discussed above) where the call-back functions
only send messages to the GUI server. For instance, if the
user selects an analysis $a$, the message ``\pr{SetAna $a$}''
is sent. The GUI server is also responsible to compute
the analysis results when they are requested by the GUI.
More details about this technique to keep states in GUIs
can be found in \cite{Hanus00PADL}. Moreover,
\cite{HanusHuchNiederau01IFL} contains a general
description of the object-oriented functional logic
programming style applied here.

It is interesting to note that the separation of the implementation
into the GUI itself and the GUI server provides also the
possibility to move the computation-intensive tasks
to a powerful ``compute server'' with only minor changes
in our implementation. Since a port can be also made externally accessible
(by changing the method to create a port, i.e.,
replacing \pr{openPort} by \pr{openNamedPort},
see Section~\ref{sec-intro-curry}),
we can start the GUI server on another machine than the GUI itself.
This requires only a change of two program lines
(replacing internal by external ports) in our implementation
and might be reasonable for more complex program analyses.

\subsection{Graphical Tracer}
\label{sec-debugger-impl}

\begin{figure}[t]
\begin{center}
  \epsfig{file=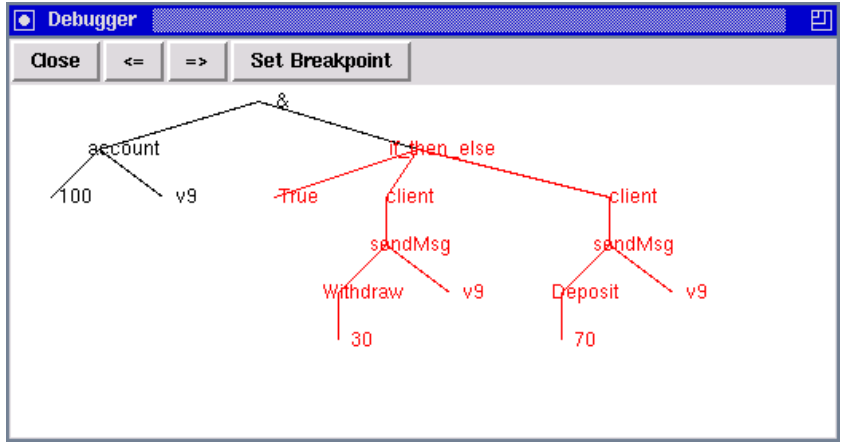,scale=0.6}
\end{center}\vspace{-3ex}
\caption{Visualization of a (concurrent) computation\label{fig-debugger}}
\end{figure}

As mentioned at the beginning, CIDER also contains a graphical debugger/tracer
to visualize the evaluation of expressions.
The debugger always shows the
expressions as trees although some parts of the expressions
are actually shared. If a subexpression is reduced,
all identical subexpressions shared with this subexpression
are also reduced in the
same step in order to be conform with Curry's operational semantics
(which is based on sharing to support laziness,
cf.\ \cite{AntoyEchahedHanus00JACM}).
The subexpression reduced in the next step (the next \pr{redex})
is always colored in red. Similarly, a variable is colored
in red if it will be bound in the next step.
This is quite useful to visualize the execution of concurrent
computations which are synchronized in Curry by the instantiation
of logical variables (compare Section~\ref{sec-intro-curry}).
One can trace forward and backward through all evaluation
steps. Furthermore, one can also set a breakpoint to skip
uninteresting parts of a computation.
A snapshot of the debugger is shown in the Fig.~\ref{fig-debugger}.

The debugger is implemented as a meta-interpreter for (Flat)Curry
in Curry, i.e., it is based on a ``step function'' that maps
a single (FlatCurry) expression into a list of expressions
according to the operational semantics of Curry as defined in
\cite{Hanus97POPL,Hanus00Curry}.
The resulting list of expressions collects all ``don't know''
alternatives of a single evaluation step, e.g.,
for purely functional programs this list contains at most one element.
Since the debugger works on the level of FlatCurry expressions,
it can visualize all kinds of computations including concurrency,
higher-order functions, etc. As a drawback, it has only a limited
performance but, nevertheless, it is useful to explain and understand
the computation model of Curry. Thus, the debugger is intended
as a teaching tool to visualize the operational semantics of Curry
rather than a tool to locate bugs in larger programs.
For the latter purpose, it might be interesting to
develop more efficient debuggers
as done for purely lazy functional languages
(e.g., \cite{ChitilRuncimanWallace01}).
The visualization part of our debugger is also implemented
in Curry by the use of the \pr{Tk} library.

\section{Extending the Development Environment}
\label{sec-ext}

This section discusses how one can extend the current
development environment in various directions.
We already mentioned that CIDER should integrate
various analysis tools for declarative programs.
In order to add a new analysis, one has to implement it
as a function of type \pr{ProgAnalysis}
(see Section~\ref{sec-ana-impl}).
For the simple addition of new analysis tools,
our implementation has a configuration module which contains
the definition of a constant
\startprog
anaList :: [(String,ProgAnalysis)]
\stopprog
This constant specifies the list of all currently available
analyses, i.e., the first component of each entry is the name
of the analysis (shown in the right column of the main window)
and the second component is the implementation of the analysis
as a function as described above. For instance, the current
definition of \pr{anaList} contains an element
\startprog
("Overlapping Rules",analyseOverlappings)
\stopprog
(compare Section~\ref{sec-ana-impl}).
Therefore, in order to integrate a new analysis into CIDER,
one only needs to add a new element to this list, recompile
the CIDER system, and the new analysis is easily accessible
through the graphical interface.

\begin{figure}[t]
\begin{center}
  \epsfig{file=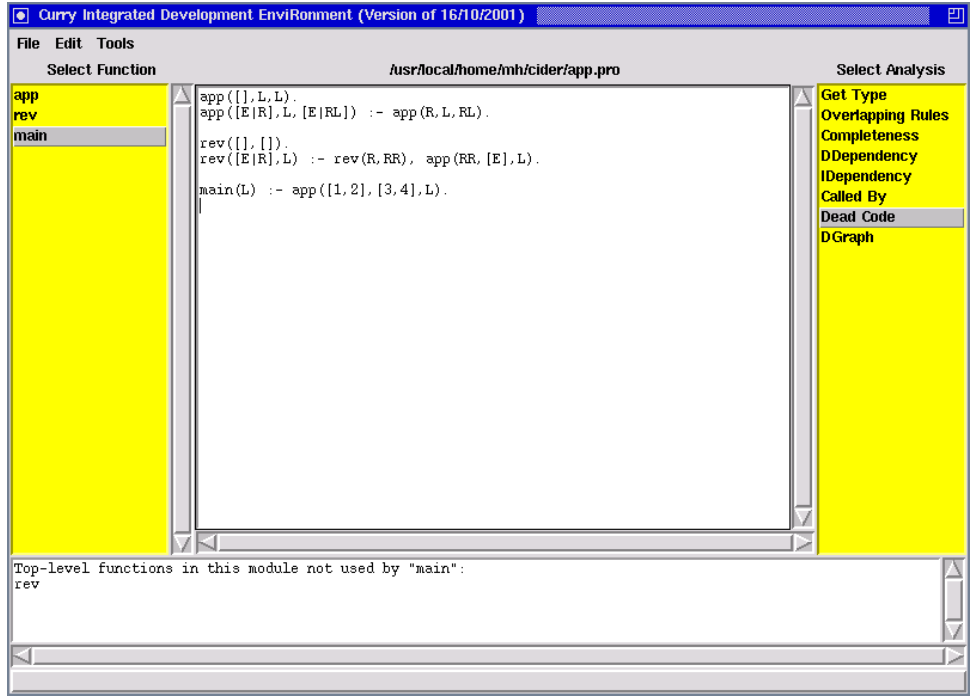,scale=0.45}
\end{center}\vspace{-3ex}
\caption{CIDER adapted to the source language Prolog\label{fig-cider-prolog}}
\end{figure}

Another possible extension of CIDER is the language of programs
to be loaded and analyzed. As explained above, CIDER is based
on the FlatCurry representation of programs which can be also
used for other source languages (e.g., Prolog, Haskell,
or Toy \cite{Lopez-FraguasSanchez-Hernandez99}). Since the program editor
in the main window is just a standard text editor, CIDER
is largely independent of the concrete syntax of the source
language. Thus, in order to adapt CIDER to another source language $\Xc$,
one only needs to replace the Curry front end (which is used
in the GUI server to load source programs) by another front end
that translates $\Xc$ programs into the corresponding
FlatCurry representation.\footnote{If the source language $\Xc$
also requires a different operational semantics of FlatCurry,
then the implementation of the tracer (see Section~\ref{sec-debugger-impl})
must be changed, too.}
Thus, the configuration module of
CIDER contains also the definition of the actual preprocessor
that translates source programs into the corresponding
FlatCurry representation. For instance, we have implemented
such a preprocessor for pure Prolog programs and
Fig.~\ref{fig-cider-prolog} shows the main window of CIDER
after loading and analyzing a Prolog program with this preprocessor.

\section{Conclusions}
\label{sec-concl}

We have presented CIDER, a graphical programming and development
environment for declarative multi-paradigm programs.
The main motivation of CIDER is to support the construction
and debugging of programs by offering a wide range of analysis
tools in an integrated manner. Although the current implementation
offers only a few basic analysis tools and is targeted at the language Curry,
we have also shown that it is fairly easy to extend CIDER
with new analysis tools or adapt it to a new declarative source language.
As far as we know, CIDER is the first development environment
for declarative multi-paradigm programs designed to integrate
various analysis tools.
The mostly related system is ${\cal IDE}$ \cite{DiosCastroGonzalezMoreno00},
a graphical development environment for the functional logic
languages Toy and Curry.
${\cal IDE}$ supports the writing of programs in a standard
text editor window and the compilation and execution of programs.
However, ${\cal IDE}$ does not offer tools for analyzing and
debugging programs.

CIDER is completely implemented in Curry and we have
discussed how advanced programming techniques are
exploited in this implementation. For instance, functions
as first-class citizens are useful to apply higher-order programming
techniques in the implementation of analysis tools or to integrate
such tools as functions in data structures. Laziness is practical
to define a program analysis in a conceptual clean manner
but compute only those parts that are actually required by the user.
Furthermore, concurrent programming is useful to structure
and distribute the various tasks possibly on different machines.

The code size of the complete implementation of CIDER is
approximately 1400 lines of Curry code. This includes
the implementation of the graphical user interface,
the various analysis tools, and the meta-interpreter
and graphical tracer. In addition, the total size
of all imported system libraries is approximately 1500 lines of Curry code.
These numbers indicate the advantage of the use of
declarative high-level programming languages
in the implementation of complex systems.

The implementation of CIDER is freely available from the web page
\verb+http://www.informatik.uni-kiel.de/~pakcs/cider/+
and requires only an installed Curry system with the libraries for
application programming distributed with the PAKCS implementation
\cite{Hanus00PAKCS}.

For future work we intend to add more analysis tools,
in particular, advanced type-based analysis tools
which can be used to analyze particular
properties of multi-paradigm programs \cite{HanusSteiner00PPDP}.
Furthermore, better (declarative) debugging tools
should be integrated into CIDER as well as
the direct efficient execution of programs by connecting
compilers based on FlatCurry, like \cite{AntoyHanusMasseySteiner01PPDP}.

\end{document}